# Trainable Neuromorphic Spintronic Hardware Via Analog Finite-Difference Gradient Methods

Catarina Pereira[1,2], Alex Jenkins[2], Eleonora Raimondo[4], Mario Carpentieri[3], Ensieh Iranmehr[2], Luana Benetti[2], Subhajit Roy[2], Ricardo Ferreira[2], Joao Ventura[1], Giovanni Finocchio[4,5], and Davi Rodrigues[3]

[1] IFIMUP, Department of Physics and Astronomy, Faculty of Sciences, University of Porto, 4169–007 Porto, Portugal

[2] International Iberian Nanotechnology Laboratory (INL), 4715-31 Braga, Portugal

[3] Department of Electrical and Information Engineering, Politecnico di Bari, 70126 Bari, Italy

[4] Istituto Nazionale di Geofisica e Vulcanologia, 00143 Rome, Italy

[5] Department of Mathematical and Computer Sciences, Physical Sciences and Earth Sciences, University of Messina, 98166, Messina, Italy

**Abstract**

Spintronic nano-neurons offer a promising route towards energy-efficient, high-performance hardware neural networks thanks to their inherent low-input nonlinear dynamics. However, training such networks remains a major bottleneck as it depends on oversimplified models of device behaviour and is highly sensitive to device variability. Here, we introduce a hardware architecture that overcomes these limitations by enabling on-device generation of gradients. First, we introduce theoretically and demonstrate experimentally that magnetic tunnel junctions can generate tunable and complex nonlinear responses. Building on this, we implement an analogue finite-difference approach to enable on-chip training in spintronic neural networks with one and two hidden layers. We experimentally implemented device- in-the-loop backpropagation in a magnetic tunnel junction–based neural network, achieving a classification accuracy of 93.3% despite pronounced device variability. During training, the gradients generated by the proposed analog neurons closely match the values derived numerically, without incurring in computational overhead. Via physical simulations, we also demonstrate that this approach can be scaled up to support training in deep architectures. Our results pave the way for reliable, trainable and fully analogue spintronic neural networks, opening up new possibilities for next-generation, energy-efficient artificial intelligence hardware.

*corresponding authors: davi.rodrigues@poliba.it, alex.jenkins@inl.int, giovanni.finocchio@unime.it

# INTRODUCTION

The rapid progress of artificial intelligence has exposed a fundamental incompatibility with the conventional von Neumann architecture[1–4]. Neural networks require dense interconnectivity, nonlinear transformations and memory-intensive operations. With the limitations of the von Neumann architecture, this translates into severe energy consumption and latency[5–7]. While von Neumann hardware excels at linear operations, this limitation has fueled the growth of neural networks with trillions of parameters, further straining computational resources and challenging scalability as transistor miniaturization slows and the benefits of Moore's law diminish[8,9]. Analog computing offer an efficient alternative for hardware implementation of neuromorphic computing by leveraging the intrinsic dynamics of physical devices to perform complex computations directly on chip[10,11]. Different technologies such as photonics and spintronics inherently support in-memory processing, nonlinearity, and massive parallelism, offering energy efficiency and computational throughput that can exceed digital approaches by orders of magnitude[12,13]. Experiments have shown that they naturally enable implementation of nonlinear transformations, generation of true stochastic signals, and performing large-scale matrix-vector multiplications within compact, low-power footprints[10,14–16].

Despite these advantages, training analog neuromorphic systems remains a major challenge[17–19]. Although neural network training presents different approaches, including statistical optimization and gradient-free methods[20,21], gradient-based training techniques[1,22,23] remain a central and effective framework due to their computational efficiency, scalability to high-dimensional parameter spaces, and strong convergence properties. The dynamics of analog devices are governed by complex and often nonideal dynamics, including stochasticity, drift, and substantial device variability. As a result, current training strategies are largely restricted to shallow architectures and typically rely on digital models of the physical devices [24–27]. Such models, however, struggle to capture the full richness and variability of analog behavior, leading to suboptimal performance and limited adaptability. Moreover, learning is generally confined to the algorithmic level, with weights programmed post hoc, thereby overlooking the possibility of online or adaptive training directly in hardware[25,28–30]. Until now, experimental spintronic implementations of in-hardware training have relied on either software-based gradient calculations or unsupervised schemes [22,25,31–33]. These approaches resulted in substantial computational overhead or limited efficiency. Although photonic systems have demonstrated in-hardware gradient generation[34] and analog finite-difference training[35,36], their implementations are typically restricted to approximated hyperbolic tangent activation functions and face scalability constraints. In addition, they are sensitive to thermal fluctuations and require sequential measurements with stored references, resulting in increased memory overhead and latency.

We propose and demonstrate an architecture based on magnetic tunnel junctions (MTJs) that enables on-chip training of neuromorphic systems without relying on digital models, see Fig. 1. First, we show a tunable nonlinear I-V response of MTJs working as neurons, going beyond conventional activation functions and operating with approximately continuous output signals, thereby leveraging the full expressiveness of nonlinear dynamics and achieving improved performance compared to binary neural network implementations[37]. MTJs can demonstrate continuous and nonvolatile tuning of their I–V characteristics[38,39]; for example, through memristive effects, it is possible to access multiple distinct and programmable resistance states[40]. Then, we present an architecture that fully leverages these nonlinear responses, incorporating on-device gradient computation through an analog implementation of the finite-difference method. The approach offers several key advantages: (i) it preserves the full expressiveness of device-level nonlinearities, (ii) inherently accounts for device-to-device variability, and (iii) enables training of deep analog architectures employing complex nonlinear activation functions. We experimentally demonstrate robust training of analog neural networks under substantial device mismatch, achieving strong generalization performance. Moreover, we show that the method scales effectively to deeper networks and supports richer activation functions that outperform conventional digital implementations, where, due to its compatibility with CMOS technology, analog MTJ neurons can be interconnected using mixed-signal architectures and integrated as localized nonlinear activation elements[16,41]. Thus, the results presented in this work support analog finite-difference training as a scalable and high-performance approach for end-to-end training of analog neuromorphic systems, contributing toward bridging the gap between physical substrates and modern machine learning and specifically for edge computing.

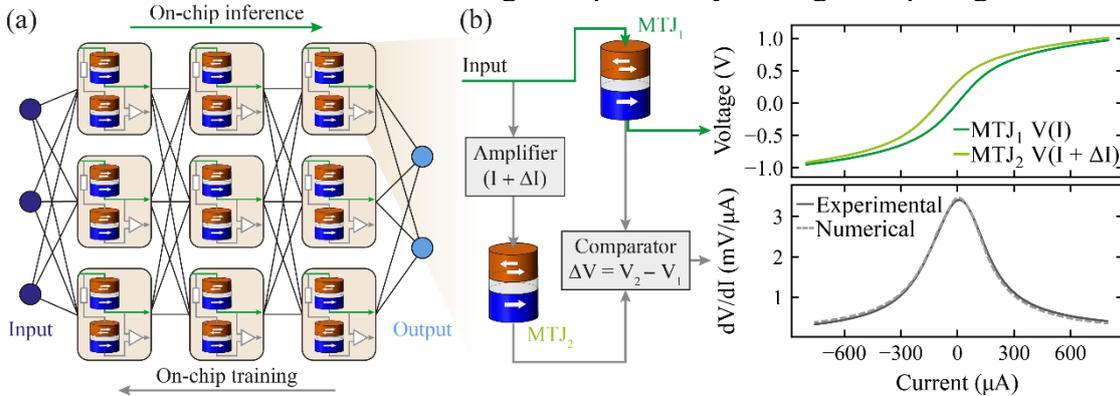

**Figure 1** - Analog finite difference method used to estimate the gradient of the MTJ current-voltage (I-V) characteristic. (a) Schematic of the proposed neuromorphic architecture enabling on-chip inference and training. (b) Neuron implementation of the analog finite-difference method using a pair of nominally identical MTJs (Stack A), simultaneously biased with input currents $I$ and $I + \Delta I$, where $\Delta I = 100$ μA. The neuron outputs both a nonlinear I-V response (green) and the associated gradient. The experimentally obtained gradients (solid lines) matches the numerically derived values (dashed lines).

## RESULTS

### TUNABLE SPINTRONIC NANO-NEURONS

MTJs are CMOS-compatible and leverage rich, tunable, nonlinear, and non-volatile magnetization dynamics to produce electrically readable signals[32], with sub-10 nm MRAM cells already demonstrated[42]. MTJs convert magnetization differences between two ferromagnetic layers into resistance changes[43,44], exhibiting a nonlinear bias dependence of tunneling magnetoresistance (TMR) arising from the relative alignment of spin-dependent densities of states [45,46]. This bias dependence varies with the parallel or antiparallel configuration of the ferromagnetic layers and can be further tuned in a non-volatile manner through local stray fields from adjacent pinned layers[47].

We investigate MTJs fabricated in two distinct configurations that present tunable nonlinear responses, referred to as Stack A and Stack B. Stack A presents a standard pinned reference layer (6 nm IrMn / 2 nm $Co_{70}Fe_{30}$ / 0.83 nm Ru / 2 nm $Co_{40}Fe_{40}B_{20}$), where the MgO insulating barrier (~1 nm) is doped with nitrogen and a free layer composed of 2.0 nm $Co_{40}Fe_{40}B_{20}$ / 0.21 nm Ta / 7 nm NiFe. In Stack B, although the pinned layer remains identical to Stack A, the MgO barrier is undoped and the free layer is modified to 2.0 nm $Co_{40}Fe_{40}B_{20}$ / 0.21 nm Ta / 3.5 nm NiFeN / 3.5 nm NiFe, introducing nitrogen specifically into the NiFe layer. The annealing process impacts the tunnel barrier profile introducing an asymmetry of the I-V curve. The I-V curves of Stack A and B are shown in Figs. 1(b) and 2, respectively.

Existing neuromorphic approaches typically use MTJs for emulation of conventional activation functions or for building blocks for reservoir computing[24,48–50]. A key result of this work is to harness the intrinsic nonlinear response of MTJs without constraining it to conventional activation function emulation. This approach enables the exploitation of richer nonlinear dynamics and intrinsic tunability. To describe the behavior of MTJ-based neurons within a layered neural network, we consider the input current to neuron ($j$) in layer ($l$) as:

$$I_j^{(l)} = \sum_i w_{ij}^{(l)} V_i^{(l-1)} + b_j^{(l)},$$

where ($w_{ij}^{(l)}$) is the synaptic weight connecting neuron ($i$) in layer ($l-1$) to neuron ($j$) in layer ($l$), ($V_i^{(l-1)}$) is the output from neuron ($i$) in the previous layer, and ($b_j^{(l)}$) is a bias term. The output of neuron ($j$) is then computed as:

$$V_j^{(l)} = f\left(I_j^{(l)}\right),$$

where $f(\cdot)$ represents the activation function realized by the MTJ device. Unlike conventional analytic activation functions such as ReLU or sigmoid[48,49], the function $f(\cdot)$ in our model is derived directly from the experimentally measured voltage response of the MTJ, without requiring any specific functional design. MTJs exhibit a rich variety of static and dynamic responses to electrically generated perturbations, including currents [24,49], electric fields [51–53], magnetic fields [16,52,54], strain [55,56], and thermal excitations[43,57,58]. This approach enables the network to naturally incorporate intrinsic device-level characteristics, such as nonlinearity, asymmetry, and saturation, into its computation.

According to the Universal Approximation Theorem, neural networks with at least one hidden layer and non-polynomial, bounded, and continuous activation functions can approximate any continuous function on a compact domain to arbitrary precision[59,60]. While many nonlinear functions have been proposed, their practical use is often limited by computational cost [61,62]. In contrast, our approach allows the experimental realization of diverse activation functions with minimal computational overhead by providing the gradient of the nonlinear function.

ANALOG FINITE DIFFERENCE METHOD

In order to address the challenge of computing gradients from device-based activation functions, we propose the analog finite-difference method, which employs a pair of nominally identical devices (MTJ$_1$ and MTJ$_2$) within a differential amplifier or comparator circuit as schematically illustrated in Fig. 1(b). In this scheme, MTJ$_1$ is driven by a reference input current $I$, while MTJ$_2$ is biased with a slightly perturbed current $I + \Delta I$. A differential amplifier or comparator circuit measures the voltage difference ($V_2 - V_1$), which estimates the gradient of the activation function. The finite-difference approximation of the gradient is then given by

$$\left.\frac{dV}{dI}\right|_{I+\frac{\Delta I}{2}} \approx \frac{V_2(I + \Delta I) - V_1(I)}{\Delta I}$$

Since the experimental gradient is obtained by applying a small current increment $\Delta I$ to one of the MTJs in each device pair, the resulting estimate corresponds to the average slope over the interval $[I, I + \Delta I]$. To align this with the ideal numerical derivative, it is necessary to account for the offset. According to the Mean Value Theorem, if the activation function is sufficiently smooth, the gradient should be evaluated at the midpoint current $I + \frac{\Delta I}{2}$. Applying this correction ensures that the experimentally measured gradient is properly centered and directly comparable to its numerical counterpart.

In Fig. 1(b), the typical I–V characteristics of Stack A are presented for $MTJ_1$ and $MTJ_2$, shown in light and dark green, respectively, and measured over a bias range of $[-800, 800]$ µA. The two devices are nominally identical, with $MTJ_2$ differing only by an additional offset current of $\Delta I = 100$ µA applied during measurement. Each MTJ is connected to an individual source meter via two probes, allowing simultaneous current application and voltage measurement. Both devices display a quasi-linear response at low bias ($-100 < I < 100$ µA), transitioning to a strongly non-linear regime at higher bias. The numerical gradient, represented in dashed lines, is compared with the experimental gradient, represented in solid lines, obtained by subtracting the outputs of the two MTJs ($\Delta V = V_2 - V_1$) and dividing by the increment ($\Delta I$). The close overlap demonstrates the accuracy and reliability of the analog finite-difference method for gradient evaluation. The radius of the MTJs under investigation were between 2.5 and 25 µm, but similar results are expected for nanometric dimensions[63] (see Figure S.2 from Supplementary Material) which have also exhibited similar non-linear I-V curve dependence. In this case, the maximum bias current varied from 500 µA up to 15000 µA, while the offset current was adjusted accordingly, spanning values from 100 to 1500 µA.

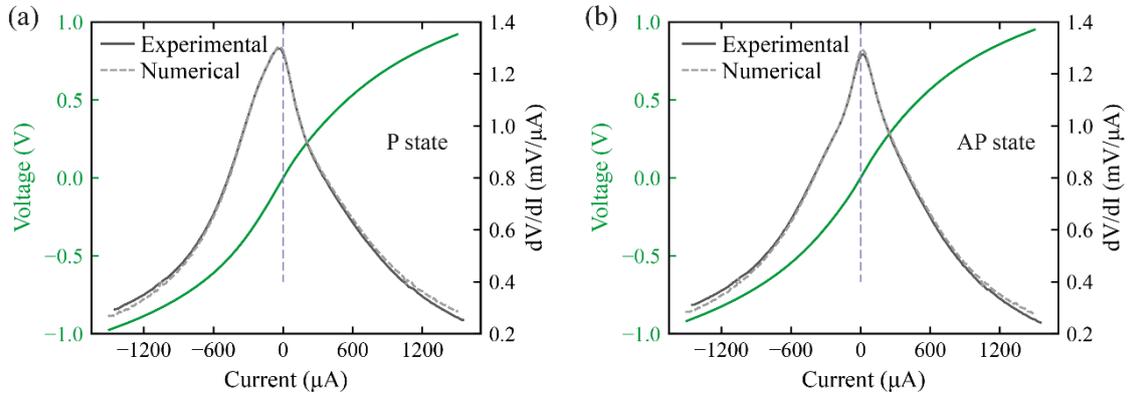

**Figure 2** - Experimentally derived activation functions and their gradients from MTJ devices. The solid curve shows the experimental gradient, while the dashed curve represents the numerically computed gradient for devices using Stack B. (a), (b) asymmetric examples highlight the range of device-level nonlinear behaviours for parallel state and antiparallel state, respectively.

Figs. 2(a)-(b) are the I-V curves measured on a device from Stack B in both the parallel (P) and anti-parallel (AP) states, respectively. In our device, the current ($I$) is the controlled input, and the voltage ($V$) is the measured output. The resulting I-V curves show an asymmetric behavior, which is more pronounced in the P state. This asymmetry arises from the different behavior of the device for positive and negative bias. Figs. 2(a)–(b) also demonstrate the tunability of MTJ activation functions via external magnetic fields, which switch the relative magnetization between P and AP states. These states exhibit distinct resistance levels through the tunneling magnetoresistance effect, yielding unique I-V characteristics and activation functions. Although the search for optimal

activation functions remains a focus of ongoing research [61,64,65], Figs. 1(b) and 2 show that the gradients obtained using the analog finite-difference method are in excellent agreement with those numerically derived from spline-fitted curves. Notably, the method remains robust and independent of the specific form of the I-V response.

The proposed analog finite-difference method offers a practical solution for extracting gradients directly from complex, hardware-specific activation responses, where conventional computational modeling often proves inadequate. A key advantage of this approach is its inherent robustness to device-to-device variability, as it relies on relative differences between paired measurements rather than requiring precise device uniformity over all neurons. This can also be extended to MTJs with substantially smaller lateral dimensions, in order of hundreds of nanometers or below, enabling operation with smaller bias currents and faster switching and readout dynamics. Consequently, MTJ readout energy lies in the picojoule range[66,67]. While the differential scheme doubles the device-level dissipation by using two MTJs, this contribution remains comparatively small, and the overall energy cost is dominated by the peripheral readout circuitry. However, MTJs are compatible with CMOS, allowing integration of low-power differential amplifiers and state-of-the-art analog-to-digital converters, with energy consumption that can be engineered to lie in the tens of picojoules depending on bandwidth, noise, and accuracy requirements[68–70].Under these conditions, the estimated energy consumption per neuron per evaluation can be on the order of approximately 150 pJ, remaining highly competitive with conventional software-based training approaches[71].

Although other approaches like sequential finite-difference on a single device reduce the number of MTJs, it requires separate time measurements, introducing temporal overhead, increased latency, and heightened sensitivity to drift and noise in physical systems. Such architecture would also require sample-and-hold or digitization circuitry, increasing complexity and energy consumption. Thus, the proposed differential approach provides instantaneous gradient estimates, enhancing robustness for in-hardware training. Although it doubles the device count, the computational cost remains independent of function complexity, which is particularly advantageous for spintronic systems that can be fabricated with minimal size and power consumption while supporting highly nonlinear responses.

NEURAL NETWORK IMPLEMENTATION

We assess the performance and reliability of the proposed architecture on two benchmark tasks using MLP architectures, including a single-hidden-layer (shallow) model and a multi-layer model. First, we consider a simple classification task based on the Iris dataset. This task was addressed through device-in-the-loop experiments, where both the I-V curves and their experimentally measured gradients were used at each training step.

In Fig. 3, we show the implementation of the Iris classification task using two architectures, with 1 hidden layer (Fig. 3(a)) and with two hidden layers (Fig. 3(b)). Each layer consisted of five neurons, and each neuron was assigned to activation and gradient characteristics derived from a distinct pair of MTJ devices. All MTJs used in this study were from Stack A. The devices are labeled as P1, P2, P3, P4, and P5 in the figures for clarity. In Fig. 3b, the second hidden layer is identical to the first. The MTJ pillar radius varied across the five pairs: three pairs had a radius of 4000 nm, one pair had a radius of 5000 nm, and one pair had a radius of 7500 nm. Fig. S.1 in the Supplemental Material shows the I-V curves and the corresponding experimental gradients for the five MTJ pairs. The training procedure was validated using both experimental and simulation approaches, in each case employing 5-fold cross-validation. Simulation results using asymmetric activation functions are presented in the supplementary material. Although the method is reliable also for asymmetric activation functions, they did not present improvement in performance.

In the experimental phase, we implemented a fully functional device-in-the-loop training process, where the real behavior of the five MTJs was used during both the forward and backward passes. At each pass, the activation functions were directly measured from the devices, while the gradients had been measured beforehand through the analog finite method and incorporated into a custom LabVIEW training environment. It is worth noting that, in the experimental setup, the gradients of the activation functions used in the program did not include the offset adjustment ($\Delta I/2$) needed to center them exactly as in the numerical gradient. The performance of the analog neural network was not affected by absence of the applied offset. For this approach, the training was performed with a single hidden layer, represented in the results by the light-blue curves.

In the simulation phase, we first measured the full I-V curves and associated measured gradients creating a Look-Up Table (LUT). Based on the LUT, we emulated the full training process. Simulations were performed using the same initial conditions as the experimental setup, i.e., identical initial weights, biases, and dataset partitions. Results were obtained for a single hidden layer (dark blue) and for a network with two hidden layers (orange).

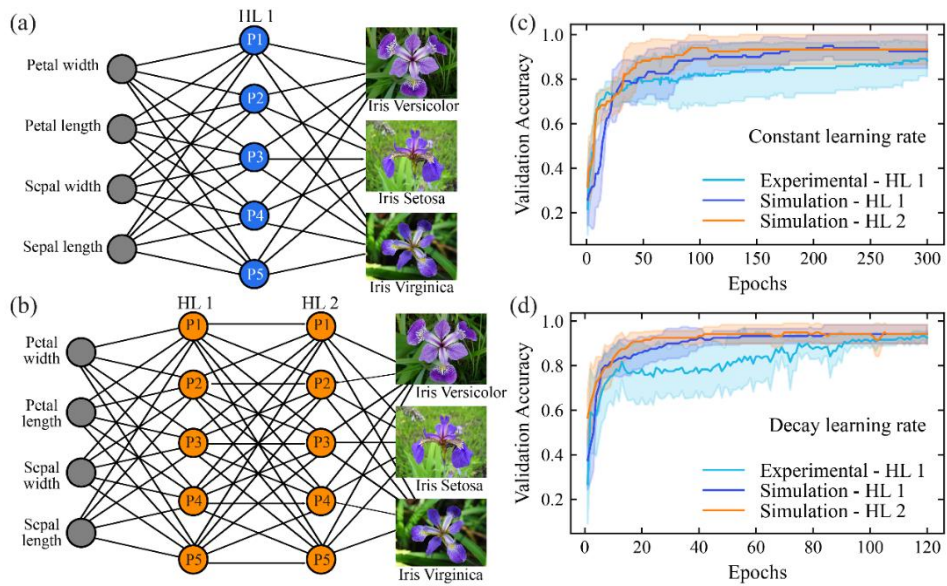

**Figure 3** - Experimental and simulated training of neural network using MTJ-based neurons for Iris classification. (a) Network architectures: a single hidden layer (HL 1) of five distinct MTJ neurons (blue), (b) Architecture with two hidden layers represented in orange (HL 1 and HL 2), where the second layer is identical to the first. (c, d) Mean validation accuracy obtained via 5-fold cross-validation. Shaded areas indicate the standard deviation across folds. Blue curves correspond to networks with one hidden layer (light = experimental, dark = simulation), while orange curves correspond to two hidden layers (simulation only). (c) Training with a constant learning rate of 0.002 over 300 epochs. (d) Training with a linearly decaying learning rate from 0.01 to 0.0088 over 120 epochs.

Fig. 3(c) and Fig. 3(d) summarize the training results for the different architectures and learning rate schedules, presenting the validation accuracy over epochs. In Fig. 3(c), training was performed with a constant learning rate of 0.002 for 300 epochs, whereas in Fig. 3(d), a linear decay schedule was applied, decreasing the learning rate from 0.01 to 0.0088 over 120 epochs.

In both cases, the training process exhibited consistent convergence with no evidence of overfitting. Accuracy increased smoothly across epochs, confirming the stability and effectiveness of the learning procedure. The linear decay schedule systematically achieved higher performance across all approaches. In the experimental implementation, the linear decay schedule reached a validation accuracy of 93.3%, compared to 89.2% for the constant learning rate.

In simulations with one hidden layer, higher performance was obtained compared to the experimental implementation, achieving 94.2% validation accuracy for the linear decay rate and 92.5% for the constant rate case. Despite this discrepancy, likely due to nonidealities in device measurements, this shows that the experimental results indicate a well-behaved training process when realized with fabricated MTJs, the network effectively learns the underlying data distribution. The observed trends closely match those from the simulations, confirming the stability and robustness of the approach.

Finally, simulations were extended to a network with two identical hidden layers. This deeper architecture achieved 94.2% validation accuracy with a constant learning rate and 95.0% with the linear decay schedule, demonstrating good convergence and outperforming the single-layer case. Unlike previous approaches, which were often limited to a single nonlinear hidden layer due to the difficulty of analytically modeling device-specific gradients [28], our results show that multilayer training with device-generated nonlinearities is both feasible and effective.

To assess scalability, we further evaluate the architecture on the MNIST handwritten digit classification task using a four-layer neural network, on a simulation level. This is critical to demonstrate that the gradient accuracy achieved by the analog method is sufficient to support backpropagation through multiple layers, even in the context of a more complex task.

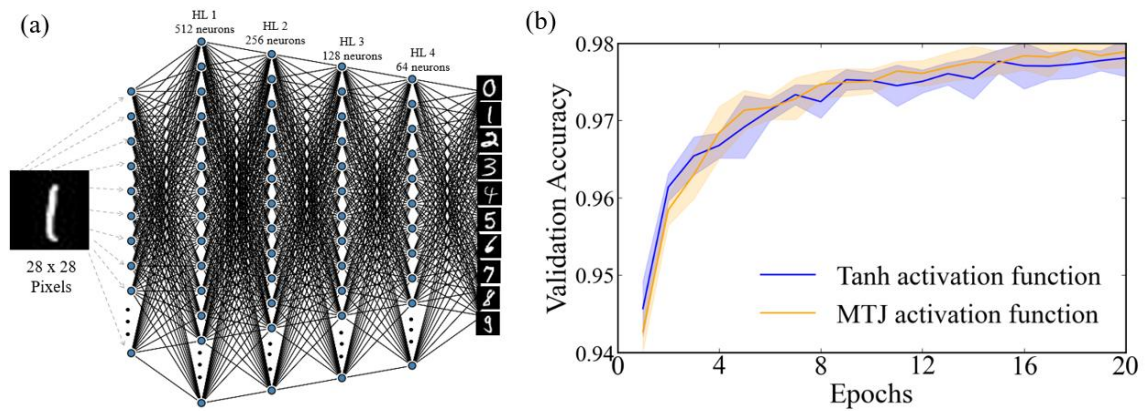

**Figure 4** - Training of a neural network using MTJ-based neurons and tanh activation for MNIST classification. (a) Network architecture: four hidden layers with 521, 256, 128 and 64 neurons, respectively. (b) Mean validation accuracy obtained via 5-fold cross-validation. Shaded areas indicate the standard deviation across folds. The orange curves represent results obtained with 15 different MTJ neurons randomly distributed across the two layers, while the blue curves show results using the standard tanh activation function.

Fig. 4(a) illustrates the architecture of the network used in this study, which has four hidden layers: the first with 512 neurons, the second with 256 neurons, the third with 128, and the fourth with 64 neurons. As in previous cases, the simulations were performed using activation functions and gradients extracted from previously characterized pairs of distinct devices. In this case, the training was conducted with 15 different devices randomly assigned across the four layers. This strategy was not implemented experimentally due to limitations of the setup but could be realized in a future system by co-integrating the MTJs with customized CMOS circuitry.

Fig. 4(b) presents a comparison between simulation results obtained using the experimental MTJ-based activation functions, shown in orange, and using a conventional hyperbolic tangent (tanh) activation function, shown in blue. These results show the validation accuracy over 20 epochs, where the final validation accuracy achieved were 97.8% for the MTJ-based system and 97.9% for the conventional digital neural network. The MTJ-based neurons model achieves slightly higher accuracy (by less than 0.1%) indicating that the MTJ-based neurons can closely match the performance of conventional functions, even when applied to more complex tasks.

The proposed architecture is particularly well-suited to energy- and area-constrained edge computing environments, where local training is essential for device adaptation and data privacy[16,72]. In such scenarios, knowledge distillation provides an efficient framework in which a student network, deployed on hardware, learns from a larger pre-trained teacher model. Thus, rather than being trained solely using the hard labels of the dataset, the student network is also trained to reproduce the teacher's output responses through soft targets, which correspond to the probability distributions produced by the teacher network. This process enables the use of a more compact student network suitable for hardware implementation, with fewer parameters than the teacher model, thereby allowing effective knowledge transfer while significantly reducing computational and memory demands[73]. To evaluate our architecture's performance under this paradigm, we used a state-of-art convolutional neural network (ResNet-18)[74] as a teacher network to guide the training of a two-layer physical neural network implemented via our analogue gradient-based approach. Using this setup, we achieved an accuracy of 97.2% on the MNIST benchmark, demonstrating that the proposed spintronic hardware architecture can efficiently support on-device learning through knowledge distillation. For details see supplemental material.

Table 1 summarizes classification performance across two benchmark tasks, Iris and MNIST, using three different approaches: experimental implementations based on MTJs, simulations using MTJ-derived activation functions, and simulations employing the standard hyperbolic tangent function. Both the experimental and simulated MTJ-based implementations achieve high classification accuracy, comparable to those obtained with conventional activation functions. Notably, the experimentally realized activation function does not lead to any degradation in performance when compared to the ideal hyperbolic tangent. The analog finite-difference gradient method is also effective in multilayer architectures, demonstrating its suitability for deeper neural networks. These results establish a promising foundation for using physically realized nonlinear functions in neuromorphic systems and enable the exploration of more complex activation functions without computational overhead through efficient gradient-based learning.

**Table 1** - Final accuracy achieved by the different neural network models.

| Dataset | Model | Network Structure | Accuracy Const. LR | Accuracy Decay LR |
|---|---|---|---|---|
| Iris | Exp. – MTJ-neurons | 4-5-3 | 89.2% | 93.3% |
| Iris | Simul. – MTJ-neurons | 4-5-3 | 92.5% | 94.2% |
| Iris | Simul. – MTJ-neurons | 4-5-5-3 | 94.2% | 95.0% |
| MNIST | Simul. – MTJ-neurons | 784-512-256-128-64-10 | 97.9% | – |
| MNIST | Simul. – `tanh` ce | 784-512-256-128-64-10 | 97.8% | – |
| Knowledge Distillation: MNIST | Teacher: ResNet-18 | Conv.: 64–128–256–512–10 | 99.1% | – |
| Knowledge Distillation: MNIST | Student: MTJ-neurons | 784-128-64-10 | 97.2% | – |

## DISCUSSION

We introduced a trainable neuromorphic spintronic hardware architecture based on an analog finite-difference gradient method and experimentally demonstrated device-in-the-loop learning. This approach offers several key advantages that make it a compelling candidate for energy-efficient and scalable neural computing. First, it preserves the full expressiveness of device-level nonlinearities, allowing complex activation functions to be directly implemented in hardware without simplification. Second, it inherently accounts for device-to-device variability through the use of paired measurements, which significantly reduces the impact of fabrication-induced mismatch and lowers the need for extensive calibration. Third, it significantly reduces the computational cost associated with nonlinear activation functions by substituting it with a linear cost in area and energy consumption. This cost structure can be further optimized through continued advances in spintronic device fabrication and integration. Fourth, the architecture supports both shallow and deep neural network models, demonstrating its flexibility and robustness across different learning tasks.

We emphasize that gradient calculation is crucial for most training algorithms[1,22], including backpropagation[75], equilibrium propagation[76,77], contrastive Hebbian learning[78], direct feedback alignment[79], and other physically-inspired training algorithms [80–82]. Thus, the analog finite-difference method introduced here provides a versatile and general framework for enabling gradient-based training directly in analog systems[83,84].

Our experimental and numerical evaluations confirmed the reliability and performance of the proposed system. In particular, the experimental realization achieved 93.3% validation accuracy for the Iris dataset. Physical simulations employing experimentally derived curves achieved 95.0% validation accuracy on the Iris dataset and 97.9% on the MNIST dataset, approaching the performance of standard digital models despite the inherent nonidealities of analog hardware. The proposed architecture is particularly well-suited to edge computing, where neural networks need to be compact and efficient to enable on-device learning, overcoming limitations of reservoir computing approaches [31,85] through direct parameter training instead of fixed dynamics with linear readout adaptation. Based on this paradigm, physical simulations employing knowledge distillation achieved 97.2% accuracy. The method proved robust even in the presence of gradient approximation errors and device variability, highlighting its suitability for practical deployment. Moreover, the proposed architecture reduces the need for repeated digital evaluation of activation functions, thereby reducing memory accesses and data movement, which are among the dominant sources of energy and latency in large-scale neural accelerators[86]. Thus, the proposed work demonstrates that the intrinsic nonlinearity of spintronic devices paves the way for more compact neural networks, as the enhanced expressiveness of nonlinear activation functions reduces the number of neurons needed to accomplish a given task.

**METHODS**
SPINTRONIC DEVICES FABRICATION

The spintronic devices investigated in this work were magnetic tunnel junctions (MTJs) fabricated with two different stacks. Stack A consisted of: [5 Ta / 50 CuN]×2 / 5 Ta / 5 Ru / 6 IrMn / 2 $CoFe_{30}$ / 0.825 Ru / / 2.6 $CoFe_{40}B_{20}$ / MgO+MgO:N / 2.0 $CoFe_{40}B_{20}$ / 0.21 Ta / 7 NiFe / 10 Ta / 10 Ru, while Stack B consisted of: 5 Ta / (50 CuN)×2 / 5 Ta / 5 Ru / 6 IrMn / 2 $CoFe_{30}$ / 0.825 Ru / 2.6 $CoFe_{40}B_{20}$ / MgO / 2.0 $CoFe_{40}B_{20}$ / 0.21 Ta / 3.5 NiFeN / 3.5 NiFe / 10 Ta / 7 Ru. All thicknesses are given in nanometers.

The samples were deposited using magnetron sputtering. For Stack A, the nitrogen-doped MgO layer was deposited in two sequential stages. In the first stage, MgO was deposited over four passes at a substrate speed of 160 mm/s under an argon (Ar) flow of 600 sccm. In the second stage, four additional passes were performed at the same speed, with a modified gas mixture consisting of 420 sccm of Ar and 200 sccm of nitrogen (N) to achieve nitrogen doping.

Device patterning was performed using direct laser writing, producing circular pillars with a radius ranging from 2500 nm to 25 000 nm. After fabrication, both configurations were annealed for 2 h under a magnetic field of 1 T. Stack A was annealed at 200 °C. For Stack B, separate sets of samples were annealed at 200 °C, 250 °C, and 330 °C.

EXPERIMENTAL NEURAL NETWORK IMPLEMENTATION

The neural networks were implemented by incorporating five distinct physical devices as artificial neurons. Each device was individually connected to a source meter. The network architecture consisted of one hidden layer comprising these five devices. Training process was performed in real time using a backpropagation algorithm developed and executed in a custom LabVIEW program. The activation function outputs were measured directly from the devices during operation. Meanwhile, the gradients for each device had been experimentally characterized beforehand and were integrated into the LabVIEW program to calculate the backpropagation algorithm. Neuron inputs were generated in software and applied to the devices, and the resulting outputs were measured directly from the five MTJs, and fed back into the software, creating a device-in-the-loop system. To improve the reliability and generalizability of the results, cross-validation was applied during training on the Iris dataset. Specifically, 20% of the data was reserved as a test set, while the remaining 80% was used for training and validation through a 5-fold cross-validation scheme. The training process was conducted under two different conditions: a fixed learning rate of 0.002 over 300 epochs, and a linearly decaying learning rate over 120 epochs. Deviations between experimental and numerical results can arise from measurement effects, including voltage readout limitations and thermal fluctuations induced by repeated measurements. These sources of variability can be mitigated by incorporating low-stress reference measurements to monitor slow drifts and applying noise-reduction strategies such as signal averaging and optimized sampling protocols.

**ACKNOWLEDGEMENTS**

This work has received funding from the European Union's Horizon research and innovation programme under grant agreement No. 101017098 (project RadioSpin), No. 101070287 (project Swan-on-chip) and No. 101070290 (NIMFEIA). The authors also acknowledge support from the Cooperation in Science and Technology action (CA23136 CHIROMAG). MC, GF and DR are with Petaspin team and thank the support of the PETASPIN association (www.petaspin.com).